\documentclass[
reprint,
superscriptaddress,
amsmath,
amssymb,
aps,
longbibliography,
pra, 
showpacs,
floatfix
]{revtex4-1}

\usepackage{color}
\usepackage{graphicx}
\usepackage{dcolumn}
\usepackage{comment}
\usepackage{array}
\usepackage{bm}
\usepackage[normalem]{ulem}

\usepackage{tabularx}
\usepackage{multirow}
\newcolumntype{d}[1]{D{.}{.}{#1}}
\newcolumntype{K}[1]{>{\centering\arraybackslash}p{#1}}
\newcolumntype{L}[1]{>{\raggedright\arraybackslash}p{#1}}
\newcolumntype{C}[1]{>{\centering\arraybackslash}p{#1}}
\newcolumntype{G}[1]{>{\raggedleft\arraybackslash}p{#1}}
\usepackage{booktabs}

\usepackage[dvipsnames]{xcolor}

\begin{document}

\title{Cavity control of nonlinear phononics}

\author{Dominik\ M.\ Juraschek}
\email{djuraschek@seas.harvard.edu}
\affiliation{Harvard John A. Paulson School of Engineering and Applied Sciences, Harvard University, Cambridge, MA 02138, USA}
\author{Tom\'{a}\v{s}\ Neuman}
\email{tomasneuman@g.harvard.edu}
\affiliation{Harvard John A. Paulson School of Engineering and Applied Sciences, Harvard University, Cambridge, MA 02138, USA}
\altaffiliation{\\ * These authors contributed equally}
\author{Johannes\ Flick}
\affiliation{Harvard John A. Paulson School of Engineering and Applied Sciences, Harvard University, Cambridge, MA 02138, USA}
\affiliation{Center for Computational Quantum Physics, Flatiron Institute, New York, NY 10010, USA}
\author{Prineha\ Narang}
\email{prineha@seas.harvard.edu}
\affiliation{Harvard John A. Paulson School of Engineering and Applied Sciences, Harvard University, Cambridge, MA 02138, USA}

\begin{abstract}
Nonlinear interactions between phonon modes govern the behavior of vibrationally highly excited solids and molecules. Here, we demonstrate theoretically that optical cavities can be used to control the redistribution of energy from a highly excited coherent infrared-active phonon state into the other vibrational degrees of freedom of the system. The hybridization of the infrared-active phonon mode with the fundamental mode of the cavity induces a polaritonic splitting that we use to tune the nonlinear interactions with other vibrational modes in and out of resonance. We show that not only can the efficiency of the redistribution of energy be enhanced or decreased, but also the underlying scattering mechanisms may be changed. This work introduces the concept of cavity control to the field of nonlinear phononics, enabling nonequilibrium quantum optical engineering of new states of matter.
\end{abstract}

\date{\today}

\maketitle


\section{Introduction}

\begin{figure}[b]
\centering
\includegraphics[scale=0.36]{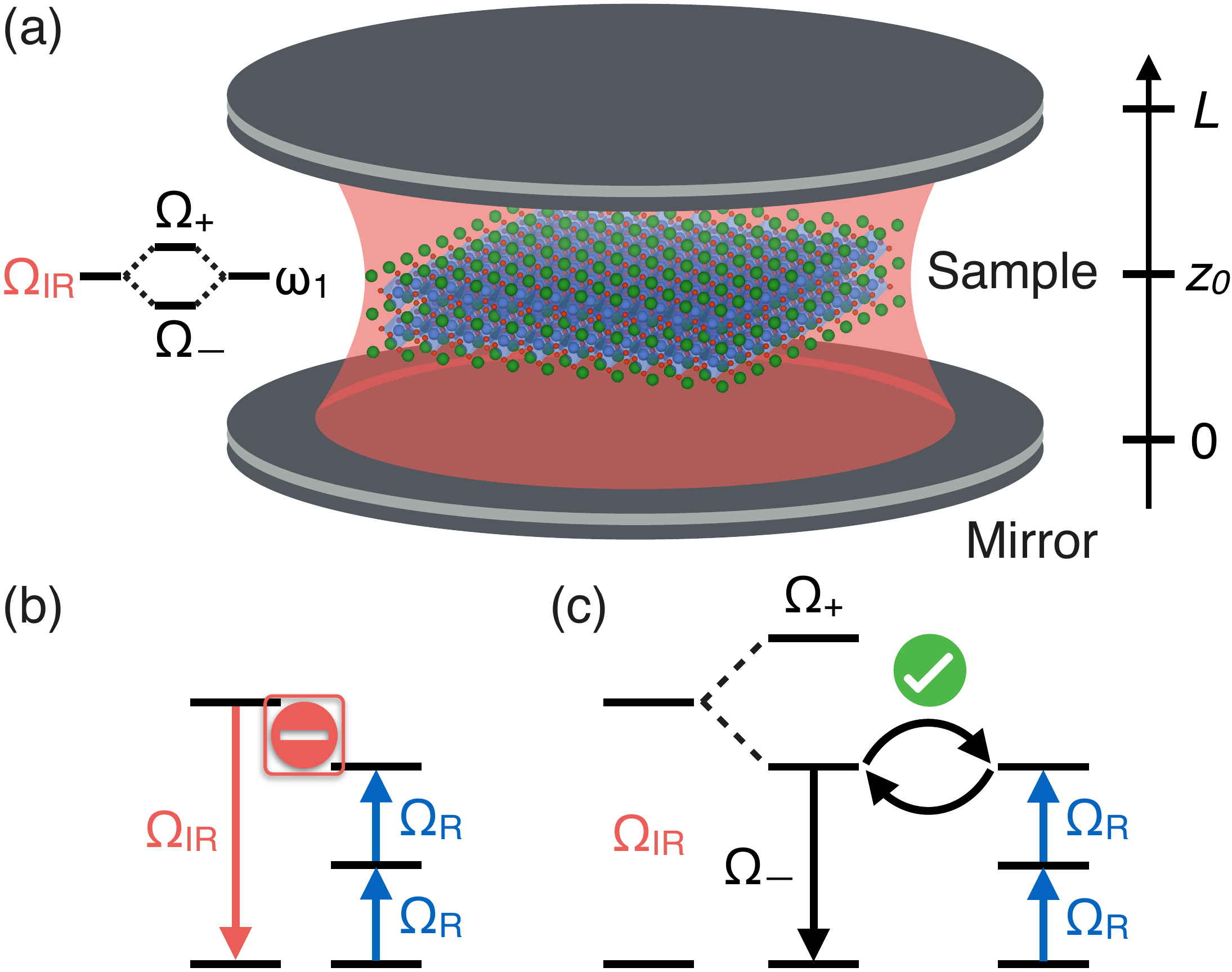}
\caption{
Cavity-mediated resonance in nonlinear phononics. (a) Schematic setup of a sample material in a terahertz cavity. The energy level diagram shows the formation of two phonon-polariton modes with frequencies $\Omega_+$ and $\Omega_-$ from the hybridization between an IR-active phonon mode, $\Omega_\mathrm{IR}$, and the fundamental cavity mode, $\omega_1$. (b) Frequency mismatch between an IR-active and two Raman-active phonon modes, $\Omega_\mathrm{IR}$ and $\Omega_\mathrm{R}$, hinders the transfer of energy between them. (c) Polaritonic splitting matches the energy of one of the branches, $\Omega_{-}$, with the Raman-active phonon modes.
}
\label{fig:cavitycoupling1}
\end{figure}

Phonon modes describe the vibrational behavior of atoms that experience a harmonic increase of the potential energy when they move away from their equilibrium positions in a crystal lattice. When the dipole moment of a phonon mode is resonantly driven with the electric field component of an ultrashort laser pulse, the amplitudes of the vibrations become so large that anharmonic corrections to the potential energy are essential. The vibrational dynamics of the solid are then governed by nonlinear interactions between the phonon modes which allow for the electronic correlations in the material to be modified, leading to emergent states of matter that do not exist in equilibrium \cite{forst:2011,Basov2017}. Two of the most prominent examples of nonlinear phonon induced states include superconductivity far above the equilibrium critical temperature \cite{mankowsky:2014,Kaiser_et_al:2014,hu:2014,Mitrano2016,Hunt2016,Mankowsky:2017} and light-induced ferroelectricity and ferroelectric switching \cite{Katayama2012,Chen2016,Mankowsky_2:2017,Kozina2019,Nova2019,Li2019}. These mechanisms are based on the energy redistribution from an initially coherently excited infrared (IR)-active phonon mode to a Raman-active one that couples to the electronic system through transient modulations of the crystal structure, exchange interaction, or the electron-phonon coupling. The efficiency of the energy transfer is however often hindered due to a frequency mismatch of the phonon modes participating in the interaction, as illustrated in Fig.~\ref{fig:cavitycoupling1}(b), limiting the exploitation of nonlinear phonon induced states to a few materials in which these resonances occur naturally.


At the same time, developments in the field of strong light-matter coupling have made it possible to engineer the optical properties of materials in optical cavities. An electromagnetic mode sustained in the cavity hybridizes with optically active states of the material to form polariton states as shown schematically in Fig.~\ref{fig:cavitycoupling1}(a). This polaritonic engineering has been used to control the chemical reactivity of molecules in their ground and excited states through a modulation of the chemical landscape \cite{Hutchison2012,flick2017,Flick2018,Feist2018,Ribeiro2018, flick2018b, rivera2018, Thomas2019}. Recently, various studies for solids have suggested an influence of strong light-matter coupling on superconductivity by exciton-photon or phonon-photon coupling \cite{Laussy2010,Cotlet2016,Sentef2018,Allocca2019,Schlawin2019,Curtis2019}. Despite the tremendous progress in theoretical methods for strong coupling in molecular and solid-state systems, the influence of strong light-matter coupling on nonlinear phonon-phonon interactions has remained unexplored, motivating this work.
Strong light-matter coupling, like nonlinear phonon interactions, relies on the matching of energies to phonon resonances in solids.


\begin{table*}[t]
\centering
\bgroup
\def\arraystretch{1.3}
\caption{
Nonlinear phononic processes. From left to right: difference-frequency ionic Raman scattering, sum-frequency ionic Raman scattering, cubic-order parametric amplification, and quartic-order parametric amplification. Shown are the terms of the lattice potential expansion in the IR and Raman-active phonon amplitudes $Q_\mathrm{IR}$ and $Q_\mathrm{IR}$ in Eq.~$\ref{eq:phononpotential}$, as well as the resonance conditions for their eigenfrequencies $\Omega_\mathrm{IR}$ and $\Omega_\mathrm{R}$.
}
\begin{tabular}{c K{3.5cm} K{3.5cm} K{3.5cm} K{4cm}}
\hline\hline
  & 
\begin{minipage}{\linewidth}
\vspace{5pt}
\includegraphics[scale=0.275]{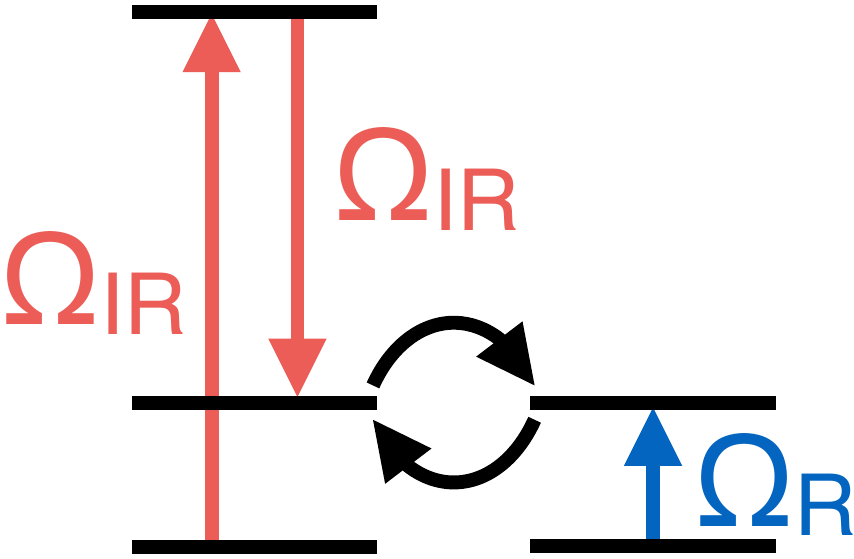}
\vspace{5pt}
\end{minipage} &
\begin{minipage}{\linewidth}
\vspace{5pt}
\includegraphics[scale=0.275]{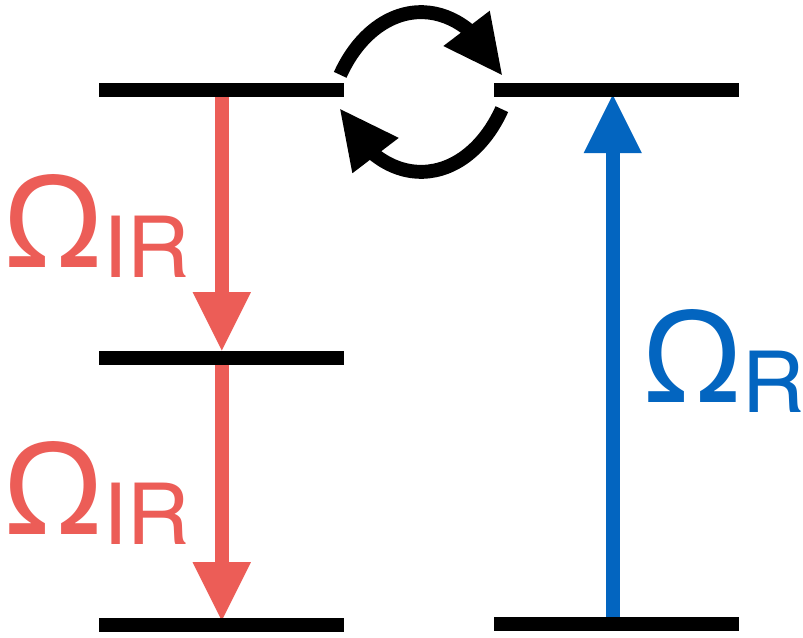}
\vspace{5pt}
\end{minipage} &
\begin{minipage}{\linewidth}
\vspace{5pt}
\includegraphics[scale=0.275]{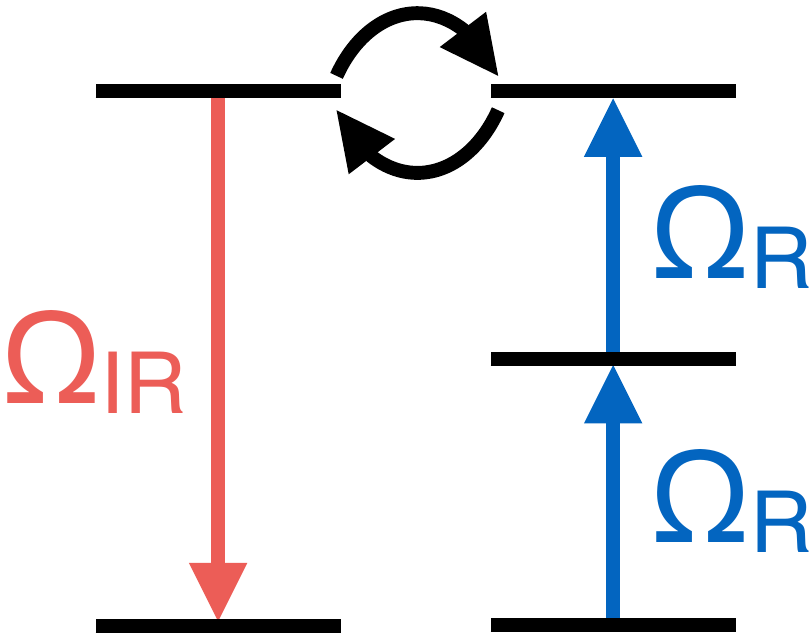}
\vspace{5pt}
\end{minipage} &
\begin{minipage}{\linewidth}
\vspace{5pt}
\includegraphics[scale=0.275]{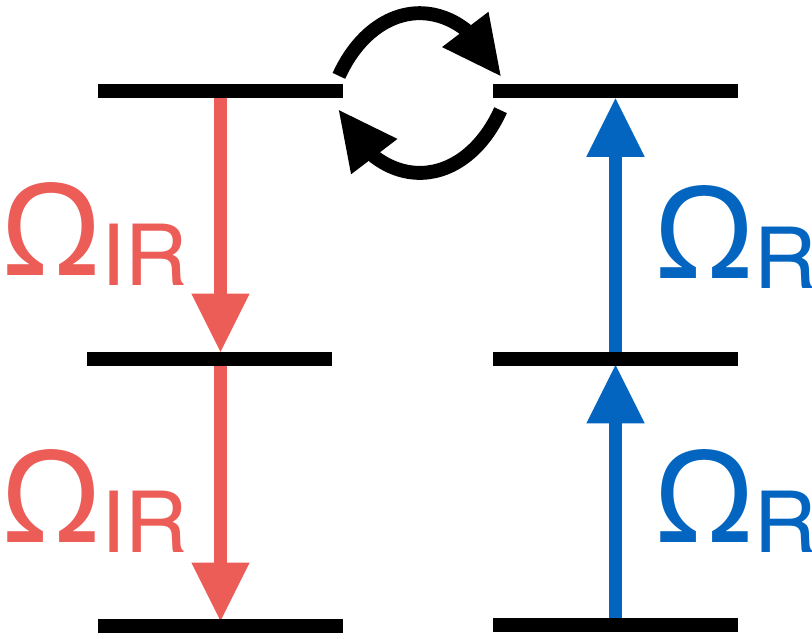}
\vspace{5pt}
\end{minipage} \\ \hline
Regime & I & II & I & III \\
Process & DF-IRS & SF-IRS & Parametric $\mathcal{O}(Q^3)$ & Parametric $\mathcal{O}(Q^4)$\\
Coupling & $Q_\mathrm{IR}^2Q_\mathrm{R}$ & $Q_\mathrm{IR}^2Q_\mathrm{R}$ & $Q_\mathrm{IR}Q_\mathrm{R}^2$ & $Q_\mathrm{IR}^2Q_\mathrm{R}^2$ \\
Resonance & Nonresonant & $\Omega_\mathrm{IR} = \Omega_\mathrm{R}/2$ & $\Omega_\mathrm{IR} = 2\Omega_\mathrm{R}$ & $\Omega_\mathrm{IR} = \Omega_\mathrm{R}\sqrt{1+e(A_\mathrm{IR}/\Omega_\mathrm{R})^2}$ \\
\hline\hline
\end{tabular}
\label{tab:orderofexcitations}
\egroup
\end{table*}

We therefore synthesize the physics of nonlinear phononics and strong light-matter coupling in this work to overcome the frequency mismatch in nonlinear phononic processes. Specifically, we demonstrate that we can tune the polaritonic splitting of an IR-active phonon mode in the cavity such that one of the polariton branches is pushed into resonance with nonlinearly coupled Raman-active phonon modes. As a result, we are able to greatly enhance the efficiency of energy redistribution through the nonlinear interaction and even change the underlying process entirely. We sketch the principle of our technique for the example of a parametric downconversion process in Figs.~\ref{fig:cavitycoupling1}(b) and (c). In Fig.~\ref{fig:cavitycoupling1}(b), an IR-active phonon mode with frequency $\Omega_\mathrm{IR}$ is out of resonance with two Raman-active phonon modes with frequencies $\Omega_\mathrm{R}$, preventing efficient coupling between them. By splitting the IR-active phonon mode in a terahertz-frequency cavity in Fig.~\ref{fig:cavitycoupling1}(c), one of the polariton branches is shifted into resonance with the Raman-active phonon modes, $\Omega_{-}=2\Omega_\mathrm{R}$, enabling the redistribution of energy between them.


\section{Nonlinear phononic processes}

We begin by analyzing the dominating nonlinear phononic mechanisms of energy redistribution from a coherently excited IR-active phonon mode to a Raman-active one, of which we provide an overview in Table~\ref{tab:orderofexcitations}. The potential energy $V$ including the most prominent phonon couplings beyond the harmonic approximation can be written as \cite{subedi:2014}
\begin{eqnarray}\label{eq:phononpotential}
V(Q) & = & \frac{\Omega_\mathrm{IR}^2}{2} Q_\mathrm{IR}^2 + \frac{\Omega_\mathrm{R}^2}{2} Q_\mathrm{R}^2 + \nonumber \\
&  & c Q_\mathrm{IR}^2 Q_\mathrm{R} + d Q_\mathrm{IR} Q_\mathrm{R}^2 + e Q_\mathrm{IR}^2 Q_\mathrm{R}^2.
\end{eqnarray}
Here, $\Omega_{\rm IR}$ and $\Omega_{\rm R}$ are the eigenfrequencies of the IR and Raman-active phonon modes in $2\pi$THz, and $Q_{\rm IR}$ and $Q_{\rm R}$ are their respective amplitudes in \AA$\sqrt{\mathrm{amu}}$, where amu is the atomic mass unit. $c$, $d$, and $e$ are the phonon coupling coefficients in meV/(\AA$\sqrt{\mathrm{amu}}$)$^n$, where $n$ is the order of expansion in the phonon amplitudes \cite{Juraschek2018}. 

Nonlinear phononic mechanisms can be classified by those linear and squared in the Raman-active phonon amplitude. In the linear case, the square of the IR-active phonon amplitude $Q_\mathrm{IR}^2$ acts as an effective driving force on the Raman-active phonon mode. If $Q_\mathrm{IR}\sim A_\mathrm{IR}\sin(\Omega_\mathrm{IR}t)$ follows a sinusoidal shape with a decaying amplitude $A_\mathrm{IR}$, then $Q_\mathrm{IR}^2\sim A_\mathrm{IR}^2[1-2\cos(2\Omega_\mathrm{IR}t)]$ possesses one static component and another oscillating at twice the frequency. Depending on whether the IR-active phonon mode has a higher or smaller frequency than the Raman-active one, this coupling leads to either difference-frequency (DF) or sum-frequency ionic Raman scattering (SF-IRS), see Table~\ref{tab:orderofexcitations}. Difference-frequency ionic Raman scattering is a nonresonant process that creates quasi-static transient distortions in the crystal structure due to its static force component on the Raman-active phonon mode, and it is held responsible for the enhancement of superconductivity in yttrium barium copper oxide \cite{mankowsky:2014,Mankowsky:2015,fechner:2016,Mankowsky:2017} and for ferroelectric switching in lithium niobate \cite{subedi:2015,Mankowsky_2:2017}. Sum-frequency ionic Raman scattering, in contrast, is a two-phonon absorption process resonant at $\Omega_\mathrm{IR}=\Omega_\mathrm{R}/2$, which has been predicted to induce Raman-active phonon mode oscillations with large amplitudes \cite{Juraschek2018,Melnikov2018,Juraschek2019_2}. In the quadratic case, the IR-active phonon amplitude acts as an effective modulation of the Raman-active phonon frequency, 
$\tilde{\Omega}_\mathrm{R}^2=\Omega_\mathrm{R}^2+2dQ_\mathrm{IR}+2eQ_\mathrm{IR}^2$.
This coupling leads to a parametric amplification of the Raman-active phonon mode in different orders, and it is resonant when either $\Omega_\mathrm{IR}=2\Omega_\mathrm{R}$ ($Q_\mathrm{IR} Q_\mathrm{R}^2$ coupling), or $\Omega_\mathrm{IR}=\Omega_\mathrm{R}\sqrt{1+e(A_\mathrm{IR}/\Omega_\mathrm{R})^2}$ ($Q_\mathrm{IR}^2 Q_\mathrm{R}^2$ coupling), see Table~\ref{tab:orderofexcitations}. Parametric amplification could be responsible for the enhancement of superconductivity through an enhancement of electron-phonon coupling \cite{Komnik2016,Knap2016,Babadi2017,Sentef2017}, and is a well-established method to generate squeezed phonon states \cite{Hu1996,Johnson2009,Fahy2016,Benatti2017}. The symmetry of the parametric coupling allows an IR-active phonon mode, in addition to Raman-active phonon modes, to also couple to silent phonon modes that are otherwise inaccessible for coherent excitation \cite{Juraschek2019_4}. In the remainder of the manuscript, we will focus on the leading-order contributions from the three-phonon processes in regimes~I and II and set $e=0$.


\section{Theoretical formalism}

In order to evaluate the interaction of the phonon modes with the cavity, we consider a quasi-1D model, in which a slab sample is placed at position $z_0$ into a cavity formed by two parabolic mirrors at positions $z=0$ and $L$, see Fig.\,\ref{fig:cavitycoupling1}(b). As only Brillouin-zone center phonon modes are relevant to the dynamics, we consider translational invariance in directions parallel to the cavity mirrors. These assumptions guarantee the transversality of the fields that enter the dynamics of the system, which simplifies their theoretical treatment.

We begin by considering Ampere's law in the Lorenz gauge expressed via the transverse vector potential $\mathbf{A}$:
\begin{align}
    - \nabla^2 \mathbf{A} + c^{-2} \partial^2_t \mathbf{A} = \mu_0 \mathbf{J}, \label{eq:weqA}
\end{align}
where $\mu_0$ is the vacuum permeability, $c$ the speed of light, and $\mathbf{J}$ is a volume current density generated by the IR-active phonon mode. $\mathbf{J}$ is related to the macroscopic polarization $\mathbf{P}$ generated by the IR-active phonon mode by
\begin{align}
\mathbf{J} = \partial_t \mathbf{P} = \frac{Z_\mathrm{IR}}{V_\mathrm{c}} \mathbf{e}_p \partial_t Q_\mathrm{IR},
\end{align}
where we have assumed the mode effective charge $Z_\mathrm{IR}$ of the IR-active phonon mode, polarized along the unit vector $\mathbf{e}_p$, to be time independent, and where $V_\mathrm{c}$ is the volume of the unit cell associated with $Z_\mathrm{IR}$. (Note that for very strong driving, the mode effective charge can be dependent on the phonon amplitude \cite{Cartella2018}.) Assuming perfectly reflecting mirrors, the vector potential of the cavity mode $m$ of polarization $p$ can be expressed as:
\begin{align}
\mathbf{A}(\mathbf{r}, t)=\sum_{m,p}\mathbf{A}_{m,p}(t)\sin (k_m z),
\end{align}
where $k_m=m\pi/L$, $m$ is a natural number, and $\mathbf{A}_{m,p}$ are time-dependent, but spatially invariant amplitudes. The two transverse polarizations $p=\{x, y\}$ of the cavity mode are degenerate. In the following, we pick the relevant polarization direction aligned with the dipole moment of the IR-active phonon mode and take into account only on the fundamental cavity mode, $m=1$, whose frequency $\omega_\mathrm{1}=\pi c/L$ is assumed to be close to $\Omega_\mathrm{IR}$. We justify this assumption a posteriori, as our results turn out to remain unchanged when we take into account higher-order cavity modes. Integrating over the spatial coordinate $z$, the equation of motion coupling the IR-active and cavity mode can be written in a scalar form, with $\mathbf{A}_{1,x} \rightarrow A_1$, as
\begin{align}
    (\omega_\mathrm{1}^2 + \partial_t) A_1 = D_1 \partial_t^2 Q_\mathrm{IR},
\end{align}
where
\begin{align}
    D_1 = \frac{2 Z_\mathrm{IR} \Delta z \sin(k_1 z_0)}{V_\mathrm{c} \varepsilon_0 L},
\end{align}
$\varepsilon_0$ is the vacuum permittivity, and $\Delta z$ is the thickness of the slab. More details of the derivation and the general form including multiple cavity modes are provided in the Supplementary Material \cite{SUPP-cavitynonlinearphononics:2019}.

Defining $G_1 \equiv \sin(k_1 z_0) Z_\mathrm{IR}$, the coupled equations of motion for the IR-active, Raman-active, and cavity modes driven by an ultrashort terahertz pulse yield
\begin{eqnarray}
\ddot{A}_1 + \kappa_1 \dot{A}_1 + \omega_1^2 A_1 & = & D_1 \dot{Q}_\mathrm{IR}, \label{eq:A1} \\
\ddot{Q}_\mathrm{IR} + \kappa_\mathrm{IR} \dot{Q}_\mathrm{IR} + \partial_{Q_\mathrm{IR}}V & = & - G_1 \dot{A}_1 + Z_\mathrm{IR} E, \label{eq:IR} \\
\ddot{Q}_\mathrm{R} + \kappa_\mathrm{R} \dot{Q}_\mathrm{R} + \partial_{Q_\mathrm{R}}V & = & 0. \label{eq:Raman}
\end{eqnarray}
Here $E(t) = E_0 \exp \{-t^2/[2(\tau/\sqrt{8\text{ln}2})^2]\} \cos(\omega_0 t)$ is the electric field component of the terahertz pulse, where $E_0$ is the peak electric field, $\omega_0$ is the center frequency, and $\tau$ is the full width at half maximum duration of the pulse \cite{Juraschek2018}. $\kappa_i$ denote phenomenological linewidths of the respective phonon and cavity modes.


\section{Nonlinear phonon dynamics}

We now investigate how the coherent nonlinear phonon dynamics change when we tune the cavity mode in and out of resonance with the IR-active phonon mode in the geometry shown in Fig.~\ref{fig:cavitycoupling1}(a). As input parameters to Eqs.~(\ref{eq:A1}), (\ref{eq:IR}), and (\ref{eq:Raman}), we choose typical values for transition-metal oxides  \cite{subedi:2015,juraschek:2017,Fechner2018,Gu2018,Khalsa2018,Juraschek2019}. We set $c=d=75$~meV/(\AA$\sqrt{\mathrm{amu}}$)$^3$, $Z_\mathrm{IR}=1$~amu$^{-1}e$, where $e$ is the elementary charge, $V_\mathrm{c}=50$~\AA$^3$, and $\kappa_i=0.05\times\Omega_i/(2\pi)$ with $i=\{1,\mathrm{IR},\mathrm{R}\}$. We find that a thickness of $\Delta z=100$~nm of the sample material is sufficient to yield significant splitting, while justifying the approximations made in our theoretical formalism, see the Supplementary Figure~S1 \cite{SUPP-cavitynonlinearphononics:2019}. In order to explore the efficiency of energy redistribution from the IR-active to the Raman-active phonon mode, we vary the frequencies of the terahertz pulse and the cavity mode, as it would be possible in an experimental setup. We look at the two different regimes, in which the resonant three-phonon processes work: Regime I, $\Omega_\mathrm{IR}>\Omega_\mathrm{R}$, for difference-frequency ionic Raman scattering and cubic-order parametric amplification, and regime II, $\Omega_\mathrm{IR}<\Omega_\mathrm{R}$, for sum-frequency ionic Raman scattering.

\begin{figure*}[t]
\centering
\includegraphics[scale=0.48]{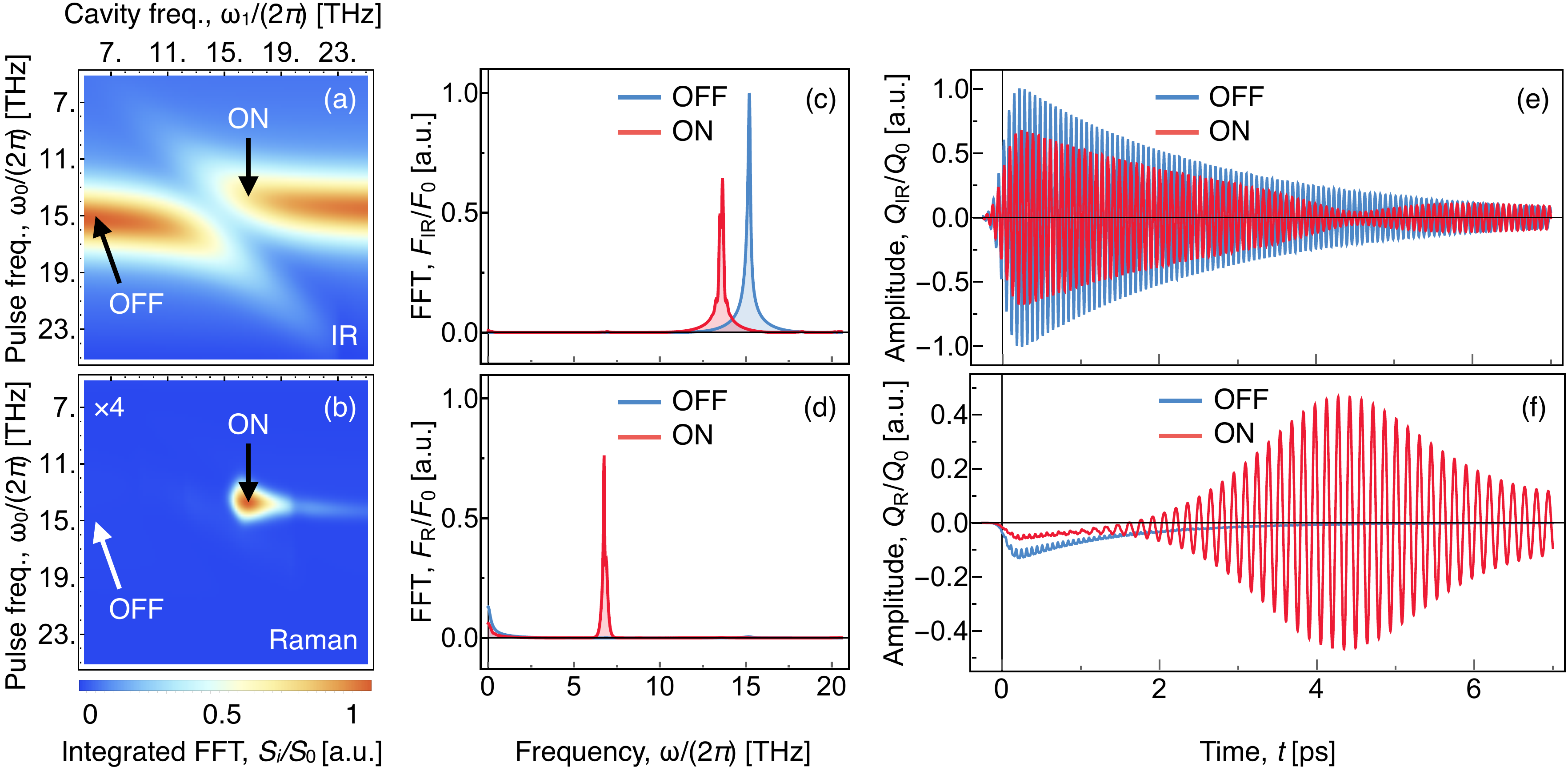}
\caption{
Coherent nonlinear phonon dynamics in regime I. (a), (b), Normalized integrated spectral densities $S_i/S_0$ of the 15~THz IR ($i$=IR) and 6.75~THz Raman-active ($i$=R) phonon modes. We vary the frequencies of the terahertz pulse, $\omega_0$, and cavity, $\omega_1$. ``OFF'' denotes a point at which the cavity mode is out of resonance with the IR-active phonon mode and there is a frequency mismatch in the parametric downconversion process, $\Omega_\mathrm{IR} > 2\Omega_\mathrm{R}$. ``ON'' denotes a point at which the cavity mode is in resonance with the IR-active phonon mode and the frequency mismatch is fixed by the lower phonon-polariton branch, $\Omega_{-} = 2\Omega_\mathrm{R}$. (c), (d), Normalized FFT spectra $F_\mathrm{IR}/F_0$ and $F_\mathrm{R}/F_0$ of the respective phonon modes when the cavity is tuned in and out of resonance with the IR-active phonon mode. (e), (f), Time evolution of the normalized phonon amplitudes $Q_\mathrm{IR}/Q_0$ and $Q_\mathrm{R}/Q_0$ of the respective phonon modes when the cavity is tuned in and out of resonance with the IR-active phonon mode.
}
\label{fig:phonondynamicsPARAMETRIC}
\end{figure*}

\subsection{Regime I:~$\Omega_\mathrm{IR}>\Omega_\mathrm{R}$}

We demonstrate our technique at the example of a 15~THz IR and a 6.75~THz Raman-active phonon mode that are detuned by $\Delta\Omega_\mathrm{IR}/(2\pi) = 1.5$~THz such that $\Omega_\mathrm{IR} - \Delta\Omega = 2 \Omega_\mathrm{R}$. We numerically solve Eqs.~(\ref{eq:A1}), (\ref{eq:IR}), and (\ref{eq:Raman}) and show the dynamics of the IR and Raman-active phonon modes in response to the excitation by a terahertz pulse with a peak electric field of $E_0=5$~MV/cm and a duration of $\tau=250$~fs in Fig.~\ref{fig:phonondynamicsPARAMETRIC}. In Figs.~\ref{fig:phonondynamicsPARAMETRIC}(a) and (b), we show maps of the integrated spectral densities, $S_i=\int_0^\infty |F_i|^2\mathrm{d}\omega$ that we obtain from Fast Fourier Transformation (FFT) spectra $F_i$ of the time-dependent phonon amplitudes of the IR ($i$=IR) and Raman-active ($i$=R) phonon modes, respectively. We plot the maps as a function of the pulse frequency $\omega_0$ (vertical axis) and cavity frequency $\omega_{1}$ (horizontal axis). The values are normalized to the maximum value that we obtain for the IR-active phonon mode, $S_0=\mathrm{max}[S_\mathrm{IR}(\omega)]$. A clear signature of phonon-photon strong coupling emerges in the spectral map of the IR-active phonon mode in Fig.~\ref{fig:phonondynamicsPARAMETRIC}(a). The maxima form two continuous branches that feature an avoided crossing as the cavity frequency $\omega_1$ is varied. When the frequency of the cavity is far detuned from the IR-active phonon frequency, $\omega_1 \gg \Omega_\mathrm{IR}$ and $\omega_1 \ll \Omega_\mathrm{IR}$, the integrated spectral density peaks when the terahertz pulse is resonant with the frequency of the IR-active phonon mode, $\omega_0=\Omega_\mathrm{IR}$. The integrated spectral density reduces as $\omega_1$ is approaching the value of $\Omega_\mathrm{IR}$ towards the center of the map, as the cavity mode hybridizes with the IR-active phonon mode and the energy is shared between the two phonon-polariton branches. In contrast, the integrated spectral density of the Raman-active phonon mode Fig.~\ref{fig:phonondynamicsPARAMETRIC}(b) exhibits a sharp peak at $\omega_1/(2\pi) \approx 16.8$~THz and $\omega_0/(2\pi) \approx 13.9$~THz. Here, the frequency of the lower phonon-polariton branch is shifted into resonance with the Raman-active phonon modes, $\Omega_{-} = 2\Omega_\mathrm{R}$, strongly enhancing the efficiency of energy redistribution. The marker ``OFF'' denotes a point at which the IR-active phonon mode is driven resonantly by the terahertz pulse, but the energy redistribution to the Raman-active phonon mode is hindered, and the marker ``ON'' denotes the resonance point of the nonlinear phononic process.

\begin{figure*}[t]
\centering
\includegraphics[scale=0.48]{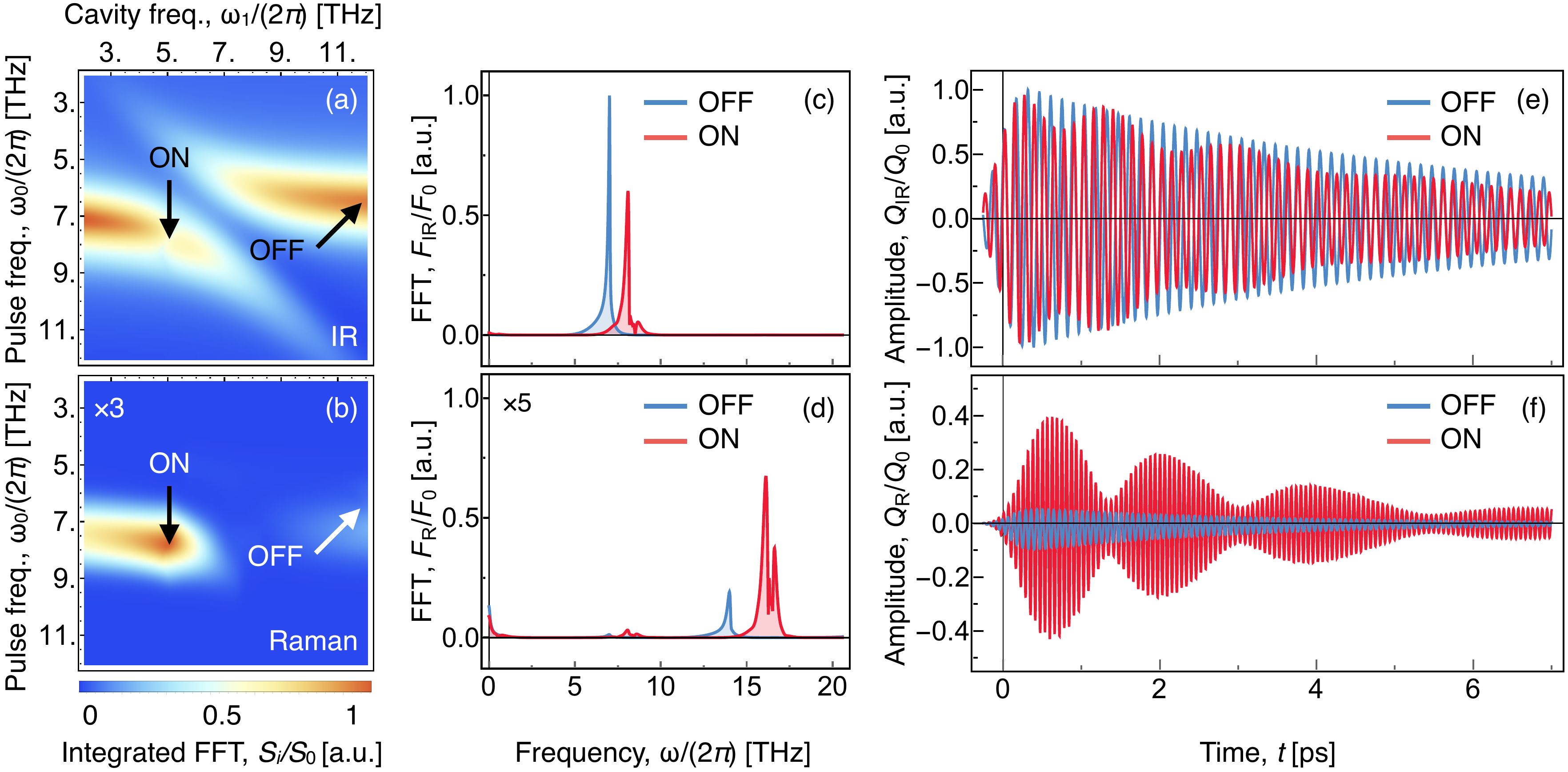}
\caption{
Coherent nonlinear phonon dynamics in regime II. (a), (b), Normalized integrated spectral densities $S_i/S_0$ of the 7.5~THz IR ($i$=IR) and 16.5~THz Raman-active ($i$=R) phonon modes. We vary the frequencies of the terahertz pulse, $\omega_0$, and cavity, $\omega_1$. ``OFF'' denotes a point at which the cavity mode is out of resonance with the IR-active phonon mode and there is a frequency mismatch in the sum-frequency ionic Raman scattering process, $2 \Omega_\mathrm{IR} < \Omega_\mathrm{R}$. ``ON'' denotes a point at which the cavity mode is in resonance with the IR-active phonon mode and the frequency mismatch is fixed by the upper phonon-polariton branch, $\Omega_{+} = \Omega_\mathrm{R}/2$. (c), (d), Normalized FFT spectra $F_\mathrm{IR}/F_0$ and $F_\mathrm{R}/F_0$ of the respective phonon modes when the cavity is tuned in and out of resonance with the IR-active phonon mode. (e), (f), Time evolution of the normalized phonon amplitudes $Q_\mathrm{IR}/Q_0$ and $Q_\mathrm{R}/Q_0$ of the respective phonon modes when the cavity is tuned in and out of resonance with the IR-active phonon mode.
}
\label{fig:phonondynamicsSF}
\end{figure*}

In order to separate the contributions from the mechanisms of difference-frequency ionic Raman scattering and parametric downconversion, we show the FFT spectra $F_\mathrm{IR}$ and $F_\mathrm{R}$ of the respective phonon modes at the points marked ``OFF'' and ''ON'' in Figs.~\ref{fig:phonondynamicsPARAMETRIC}(c) and (d). The values are normalized to the maximum value that we obtain for the IR-active phonon mode, $F_0=\mathrm{max}[F_\mathrm{IR}(\omega)]$. The FFT spectrum of the IR-active phonon mode in Fig.~\ref{fig:phonondynamicsPARAMETRIC}(c) shows a peak at its eigenfrequency of $\Omega_\mathrm{IR}/(2\pi)=15$~THz, when the cavity is out of resonance, and a peak at the frequency of the lower phonon polariton $\Omega_{-}/(2\pi)=13.5$~THz, when the cavity is tuned into resonance, thus fulfilling the resonance condition for the cubic-order parametric downconversion process, $\Omega_{-} = 2\Omega_\mathrm{R}$. The FFT spectrum of the Raman-active phonon mode in Fig.~\ref{fig:phonondynamicsPARAMETRIC}(d) shows only a static component arising from difference-frequency ionic Raman scattering when the cavity is out of resonance. When the cavity is tuned into resonance, this static component decreases, but a large peak at its eigenfrequency of $\Omega_\mathrm{R}/(2\pi)=6.75$~THz emerges.

The cavity-induced resonance becomes even more distinct when looking at the time-dependent evolutions of the phonon amplitudes $Q_\mathrm{IR}$ and $Q_\mathrm{R}$ in Figs.~\ref{fig:phonondynamicsPARAMETRIC}(e) and (f). The values are normalized to the maximum value that we obtain for the IR-active phonon mode, $Q_0=\mathrm{max}[Q_\mathrm{IR}(t)]$. The IR-active phonon mode shows the common evolution of a resonantly driven and slowly decaying phonon mode in Fig.~\ref{fig:phonondynamicsPARAMETRIC}(e) when the cavity is out of resonance. In resonance, the amplitude exhibits an additional decrease and recovery around 4~ps from the redistribution of energy to the Raman-active phonon mode. $Q_\mathrm{R}$ in turn exhibits a transient shift of its equilibrium position and only tiny oscillatory amplitude in Fig.~\ref{fig:phonondynamicsPARAMETRIC}(f) when the cavity is out of resonance, a common feature of difference-frequency ionic Raman scattering. When the cavity is tuned into resonance, the static shift decreases, but a large-amplitude oscillation builds up over time through the cubic-order parametric downconversion process. The time delay of the build-up depends on various factors, such as the relative frequencies of the phonon modes, their dampings, and the strength of their nonlinear coupling.

\subsection{Regime II:~$\Omega_\mathrm{IR}<\Omega_\mathrm{R}$}

We now investigate the case of a 7.5~THz IR and a 16.5~THz Raman-active phonon mode that are detuned by $\Delta\Omega_\mathrm{IR}/(2\pi) = 0.75$~THz such that $2(\Omega_\mathrm{IR} - \Delta\Omega_\mathrm{IR}) =  \Omega_\mathrm{R}$. We show the numerical results for an excitation of the phonon modes by a terahertz pulse with a peak electric field of $E_0=2$~MV/cm and a duration of $\tau=500$~fs in Fig.~\ref{fig:phonondynamicsSF}, which is structured in the same way as Fig.~\ref{fig:phonondynamicsPARAMETRIC}. Analog to the behavior in regime~I, the spectral map of the IR-active phonon mode in Fig.~\ref{fig:phonondynamicsSF}(a) shows the formation of phonon-polariton branches as a result of phonon-photon strong coupling. The integrated spectral density of the Raman-active phonon mode Fig.~\ref{fig:phonondynamicsSF}(b) exhibits a peak around $\omega_0/(2\pi) \approx 7.8$~THz and $\omega_1/(2\pi) \approx 5.2$~THz, where the frequency of the upper phonon-polariton branch is shifted into resonance with the Raman-active phonon mode, $\Omega_{+} = \Omega_\mathrm{R}/2$, strongly enhancing the efficiency of energy redistribution. The peaks is not as sharp as in the case of parametric downconversion, but the redistribution of energy is so large, that it visibly imprints on the spectral map of the IR-active phonon mode.

We find that the term $dQ_\mathrm{IR}Q_\mathrm{R}^2$ is negligible in this regime, as the dynamics remain unchanged when we set $d=0$. The features of the FFT spectra $F_\mathrm{IR}$ and $F_\mathrm{R}$ of the respective phonon modes in Figs.~\ref{fig:phonondynamicsSF}(c) and (d) therefore arise from sum-frequency ionic Raman scattering entirely. In this case, the frequency range of the cavity is close enough to the IR-active phonon mode in order to induce a shift even at the point marked as ``OFF'' resonant, as can be seen by a shift of the frequency in Fig.~\ref{fig:phonondynamicsSF}(c) away from the eigenfrequency $\Omega_\mathrm{IR}$. In the resonant case, ``ON'', the peak of the FFT spectrum lies around 16.2~THz, which does not yet perfectly match the eigenfrequency of the Raman-active phonon mode, $\Omega_\mathrm{R}/(2\pi)=16.5$~THz. The consequences for the Raman-active phonon mode can be seen in Fig.~\ref{fig:phonondynamicsSF}(d). ``OFF'' resonance, the Raman-active phonon mode shows a small peak at $2\Omega_\mathrm{IR}$ in addition to a tiny static component. In the resonant case, however, a large double peak emerges. This is a result of the Raman-active phonon mode simultaneously following the two-frequency component of the upper phonon polariton, $2\Omega_\mathrm{+}$, and gaining significant spectral weight at its eigenfrequency $\Omega_\mathrm{R}/(2\pi)=16.5$~THz with the same magnitude.

The IR-active phonon amplitude $Q_\mathrm{IR}$ in Fig.~\ref{fig:phonondynamicsPARAMETRIC}(e) shows the common evolution of a resonantly driven and slowly decaying phonon mode when the cavity is out of resonance, as in regime~I. With the cavity tuned into resonance, $Q_\mathrm{IR}$ exhibits a small beating signal in addition, arising from the mutual energy exchange with the Raman-active phonon mode. The Raman-active phonon amplitude $Q_\mathrm{R}$ is oscillating with small amplitude in the ``OFF''-resonant case in Fig.~\ref{fig:phonondynamicsPARAMETRIC}(f) and gets strongly enhanced when the cavity is tuned into resonance with the IR-active phonon mode (``ON''). While the beating is small compared to the amplitude of the IR-active phonon mode, it is significant for the Raman-active one. This example shows that even when the IR-active phonon mode is not perfectly shifted into resonance with the Raman-active one, the phonon amplitude can still be strongly enhanced. The increase of $Q_\mathrm{R}$ by a factor of four in this case increases the efficiency of energy redistribution by $\times 16$.


\section{Conclusion}

The ultimate goal of the technique presented here is to broaden the range of materials in which resonant nonlinear phononic processes can be exploited to yield nonequilibrium states of matter. Concepts in cavity control of nonlinear processes enable a new pathway for quantum optical engineering of new states of matter. The analysis presented here is applicable to resonant coupling mechanisms between IR-active phonon modes and other fundamental excitations in solids and molecules. In particular, we anticipate that it will be used to enhance processes involving the amplification of Goldstone excitations \cite{Buzzi2019,Juraschek2019_4}, the coupling to magnetic degrees of freedom \cite{Li2018,Sivarajah2019}, and the control of superconducting order parameters \cite{VonHoegen2019}.


\begin{acknowledgments}
We are grateful to Morgan Trassin (ETH Zurich), Denis Gole\v{z}, Antoine Georges (CCQ), and Eugene Demler (Harvard) for useful discussions. 

This work was supported by the DARPA DRINQS Program  under  award number D18AC00014 as well as the Swiss National Science Foundation (SNSF) under project ID184259. T.N. acknowledges the 'Photonics at Thermodynamic Limits' Energy Frontier Research Center funded by the U.S. Department of Energy, Office of Science, Office of Basic Energy Sciences under Award Number DE-SC0019140 that supported the computational approaches used here. J. F. acknowledges partial financial support from the Deutsche Forschungsgemeinschaft (DFG) under contract No. FL 997/1-1. P. N. is a Moore Inventor Fellow supported by the Gordon and Betty Moore Foundation and is a CIFAR Azrieli Global Scholar.
\end{acknowledgments}



%

\end{document}